\documentclass[12pt]{article}
\usepackage{amssymb}
\usepackage{amsmath}
\usepackage{hyperref}
\begin{document}
\title{\bf Shear Viscosity to Entropy Density for a Black Brane in $5$-dimensional Einstein-Yang-Mills Gravity}
\author{Mehdi Sadeghi\thanks{Email: mehdi.sadeghi@modares.ac.ir}  \hspace{2mm} and
        Shahrokh Parvizi\thanks{Corresponding author: Email: parvizi@modares.ac.ir}\hspace{1mm}\\
		{\small {\em  Department of Physics, School of Sciences,}}\\
        {\small {\em Tarbiat Modares University, P.O.Box 14155-4838, Tehran, Iran}}\\
       }
\date{\today}
\maketitle

\abstract{We calculate the ratio of shear viscosity to entropy density for a  black brane of $5$-dimensional Einstein-Yang-Mills Gravity by Kubo and membrane paradigm methods. The former gives $\frac{1}{4\pi}$ exactly which is an expected result in the context of the Einstein-Hilbert gravity. In contrast, the membrane paradigm reaches $\frac{1}{4\pi}$ only in large horizon radius limit. We comment on this discrepancy. }\\



\section{Introduction} \label{intro}

\indent AdS/CFT duality introduced by Maldacena \cite{Ref1,Ref2,Ref3,Ref4} relates two kinds of theories: gravity in (n+1)-dimension and field theory in n-dimension. The most familiar example, the AdS/CFT duality asserts that $SU(N)$ ${\mathcal N}=4\ $Super Yang-Mills (SYM) theory is dual to Type IIB string theory on $AdS_{5} \times S^{5} $. There's no way to solve the strongly coupled field theories either analytically or perturbatively. AdS/CFT duality is a technique to overcome this problem. By using this duality, we can translate the strongly coupled field theory into a weakly gravitational theory and vice versa.  In the long wavelength limit this duality leads to fluid/gravity duality \cite{Ref5,Ref6,Ref7}. Any fluid is characterized by some transport coefficients. These coefficients identify the underlying microscopic properties of fluids which in turn rooted in the field theory interactions at strong coupling. So the fluid/gravity duality would be a proper tool to calculate these coefficients. In this work, our interest is the shear viscosity, one of the transport coefficients. The conservation of energy and momentum in relativistic Hydrodynamics is as follows \cite{Ref8,Ref9},
\begin{align}
\nabla _{\mu } T^{\mu \nu } &=0, \\
T^{\mu \nu } &=(\rho +p)u^{\mu } u^{\nu } +pg^{\mu \nu }.
\end{align}

Note that the term ``relativistic fluid'' doesn't mean the fluid is necessarily moves near the speed of light. However, the Lorentz symmetry preserves in the relativistic fluid \cite{Ref9}. If there is a chemical potential in the system we have also a current conservation,
\begin{align}
\nabla _{\mu } J^{\mu } &=0, \\
J^{\mu } &=\rho \, u^{\mu }.
\end{align}

One can introduce a parameter expansion $\varepsilon =\frac{l_{mfp} }{L} $, where $l_{mfp}$ and $L$ are the mean free path and  the characterized length of system or the scale for the field fluctuations, respectively.  The scale of field variations has to be large compared to the mean free path, $l_{mfp} \ll L$,  for the validity of hydrodynamics regime. We know that the regime where the fluid is valid corresponds to a theory with large AdS black holes. We can expand the energy-momentum tensor in terms of $\varepsilon$ when it is smaller than $1$ \cite{Ref6,Ref7}.
\begin{align}
& T^{\mu \nu } =(\rho +p)u^{\mu } u^{\nu } +pg^{\mu \nu } -\sigma ^{\mu \nu }\\
&\sigma ^{\mu \nu } = {P^{\mu \alpha } P^{\nu \beta } } [\eta(\nabla _{\alpha } u_{\beta } +\nabla _{\beta } u_{\alpha })+ (\zeta-\frac{2}{3}\eta) g_{\alpha \beta } \nabla .u]\\& P^{\mu \nu }=g^{\mu \nu}+u^{\mu}u^{\nu}. \nonumber
\end{align}
\noindent where $\eta$ and $\zeta $ are shear and bulk viscosities, respectively.\\

\indent There are different methods for calculating shear viscosity such as: boosted black brane, membrane paradigm and Green-Kubo formula \cite{Ref10}. Two latter formalisms are used in this article. In the Green-Kubo formula approach, transport coefficients of plasma are related to their thermal correlators. The Green-Kubo formula derived from linear response theory \cite{Ref11,Ref12,Ref13} can be expressed as,
\begin{equation}
\eta =\mathop{\lim }\limits_{\omega \to 0} \frac{1}{2\omega } \int dt\,  d\vec{x}\, e^{i\omega t} \left\langle [T_{y}^{x} (x),T_{y}^{x} (0)]\right\rangle =-\mathop{\lim }\limits_{\omega \, \to \, 0} \frac{1}{\omega } \Im G_{y\, \, y}^{x\, \, x} (\omega ,\vec{0}).
\end{equation}

The ratio of shear viscosity to entropy density is proportional to the inverse square coupling  of quantum thermal field theory,
 \begin{equation}
\frac{\eta }{s} \sim \frac{1}{\lambda^2 }
 \end{equation}
where $ \lambda $ is the coupling constant of field theory. In particular, the stronger the coupling, the weaker the shear viscosity per entropy density. In theories with Einstein gravity duals, even in the limit of  infinite coupling the ratio  $ \frac{\eta}{s} $ saturates the bound  $ \frac{1}{4\pi} $ \cite{Ref6} and only for higher derivative gravities it may violate the bound. 

In the following, we review the black brane in the Einstein-Yang-Mills theory. Then calculate the shear viscosity to the entropy density ratio by two methods of the  Green-Kubo formula and the membrane paradigm and compare the results. While by the Green-Kubo the ratio is found to be exactly  $\frac{1}{4\pi}$, the membrane paradigm reaches this value only at large black hole limit which is the correct hydrodynamic limit of the theory. 
\section{Black Brane in 5-d Einstein-Yang-Mills Gravity}
\label{sec2}

\indent Consider the Einstein-Yang-Mills action,

\begin{equation}\label{eq08}
S=\int d^{5}  x\sqrt{-g} (R+\frac{12}{l^{2} } -\gamma _{ab} F_{\mu \nu }^{(a)} F^{(b)\, \, \mu \nu }),
\end{equation}
where $R$ is the Ricci scalar, $l$ the AdS radius and $F_{\mu \nu }^{(a)} $ the $SO(5,1) $ Yang-Mills gauge field tensor \cite{Ref14},
\begin{align} \label{eq03}
F_{\mu \nu }^{(a)} =\partial _{\mu } A_{\nu }^{(a)} -\partial _{\nu } A_{\mu }^{(a)} +\frac{1}{2e} C_{bc}^{a} A_{\mu }^{(b)} A_{\nu }^{(c)} \;\;\; a=1,\, 2,...,N,
\end{align}
in which $e$ is the gauge coupling constant, $C$'s are the gauge group structure constants and $A_{\nu }^{(a)} $'s the gauge potentials.

\noindent The metric tensor of the gauge group is,
\begin{align*}
&\gamma _{ab} =-\frac{\Gamma _{ab} }{|\det \Gamma _{ab}|^{^{\frac{1}{N} } } },\\&
\Gamma _{ab} =C_{ad}^{c} C_{bc}^{d} \, \, , \, \, \, \, \, \, \, |\det \Gamma _{ab}|> 0.
\end{align*}

Variation of the action (\ref{eq08}) with respect to the spacetime metric $g_{\mu \nu } $ and the gauge potential $A_{\nu }^{(a)} $ yields,
\begin{align}
&G_{\mu \nu } +\Lambda g_{\mu \nu } =8\pi T_{\mu \nu },\\&
F_{;\nu }^{(a)\mu \nu } =j^{(a)\mu },
\end{align}
where the gauge current and the stress-energy tensor carried by the gauge fields are as follows,
\begin{align}
&T_{\mu \nu } =\frac{1}{4\pi } \gamma _{ab} (F_{\mu }^{(a)\lambda } F_{\nu \lambda }^{(b)} -\frac{1}{4} F^{(a)\lambda \sigma } F_{\lambda \sigma }^{(b)} g_{\mu \nu } ),\\&
j^{(a)\nu } \, =\frac{1}{e} C_{bc}^{a} A_{\mu }^{(b)} F^{(c)\mu \nu }.
\end{align}

\noindent The invariant scalar ${\mathcal F}\equiv \gamma _{ab} F^{(a)\lambda \sigma } F_{\lambda \sigma }^{(b)}$ for the YM fields is,
\begin{equation}
 {\mathcal F}_{YM} =\frac{6e^{2} }{r^{4} } .
\end{equation}

\noindent So that, the solution for this action will be as below \cite{Ref14},
\begin{equation}\label{eq09}
ds^{2} =-f(r)dt^{2} +\frac{dr^{2} }{f(r)} +r^{2} d{\Omega}_{k}^2,
\end{equation}
where,
\begin{equation}
f(r)= k-\frac{3m}{r^{2} } -\frac{2e^{2} \ln r}{r^{2} } +\frac{r^{2} }{l^{2} } .
\end{equation}
When $ k=-1 $ we have topological black hole, $ d{\Omega}_{k=-1}^2=d\Sigma^2_{3} = d\theta ^{2} +\sinh^{2} \theta \, (d\varphi ^{2} +\sinh^{2} \varphi \, d\psi ^{2})$. If $ k=1 $ and for a small portion of the solid angle we have  black brane, $ d\Omega_{k=1}^{2}=\frac{1}{l^{2}}d\vec{x}^2 =\frac{1}{l^{2}}( dx_{1} ^{2} +dx_{2} ^{2} + dx_{3}^{2})$.

\indent This is a black brane with mass related to parameter $m$ and YM charge $e$. Event horizon is where $g^{rr} (r)=0$. Topology of this black brane is $R \times M ^{4} $ where $M^{4} $ is a $4$-dimensional manifold with constant positive curvature. The horizon radius $r_{+} $ is obtained as follows,
\begin{align*}
&f(r=r_{+} )=1-\frac{3m}{r_{+}^{2} } -\frac{2e^{2} \ln r_+}{r_{+}^{2} } +\frac{r_{+}^{2} }{l^{2} } =0
\end{align*}
Here  $m$ is replaced in favor of other constants,
\begin{align}
3m&=\frac{r_{+}^{4} }{l^{2} } +r_{+}^{2} -2e^{2} \ln r_{+}\\
f(r)&=1-\frac{1}{r^{2} } (\, \frac{r_{+}^{4} }{l^{2} } +r_{+}^{2} -2e^{2} \ln r_{+} )-\frac{2e^{2} \ln r}{r^{2} } +\frac{r^{2} }{l^{2} } .
\end{align}

\noindent Hawking temperature is defined by,
\begin{equation} \label{eq18}
\begin{split}
T=\frac{\kappa (r_{h} )}{2\pi } &= \frac{1}{2\pi \sqrt{g_{rr} } } \frac{d}{dr} \sqrt{g_{tt} }|_{r=r_{h} }=\frac{1}{4\pi \sqrt{g_{rr} g_{tt} } } \partial _{r} g_{tt}|_{r=r_{h} }  \\
 &=\frac{1}{4\pi } \partial _{r} g_{tt}|_{r=r_{h} } =\frac{1}{4\pi } \partial _{r} f(r)|_{r=r_{h} }
\end{split}
\end{equation}
where, $\kappa (r_{h} )$ is the surface gravity on the event horizon.
In our case temperature is,
\begin{equation} \label{eq19}
T=\frac{1}{4\pi } \partial _{r} f(r)|_{r=r_{+} } =\frac{2r_{+}^{4} +l^{2}(r_{+}^{2} - e^{2}) }{2\, \pi \, r_{+}^{3} l^{2} } .
\end{equation}
For large $ r_{+} $, the temperature is,
\begin{equation}
T=\frac{r_{+} }{\pi l^{2}} .
\end{equation}

The entropy can be found by using Hawking-Bekenstein formula  \cite{Ref15},
\begin{eqnarray}
A&=&\int d^{3} x \sqrt{-g} |_{r=r_+,t=cte}= \frac{r_{+}^{3} V_{3}}{l^{3}} \nonumber\\
S&=&\frac{A}{4G} =\frac{r_{+}^{3} V_{3} }{4l^{3}G} \nonumber\\
s&=&\frac{S}{V_{3} } =\frac{4\pi r_{+}^{3} }{l^{3}}
\end{eqnarray}
where $V_{3}$ is the volume of the constant $t$ and $r$ hyper-surface with radius $r_{+}$ and in the last line we used $\frac{1}{16\pi G} =1$ so $\frac{1}{4G} =4\pi$.

Notice $r$ is the radial coordinate that put us from bulk to boundary. In the following we apply dimensionless variable $u$ instead of r, that is $u=(\frac{r_{+} }{r} )^{2}$ , then
\begin{equation} \label{eq20}
f(r)=1-\frac{1}{r^{2} } (\, \frac{r_{+}^{4} }{l^{2} } +r_{+}^{2} -2e^{2} \ln r_{+} )-\frac{2e^{2} \ln r}{r^{2} } +\frac{r^{2} }{l^{2} } ,
\end{equation}
in terms of $u$ it found to be
\begin{equation} \label{eq22}
 f(u)=1-u-\frac{ur_{+}^{2} }{l^{2} } +\frac{r_{+}^{2} }{ul^{2} } +\frac{e^{2} \, u\, \ln u}{r_{+}^{2} } .
\end{equation}
So that,
\begin{equation} \label{eq23}
\begin{split}
ds^{2} &=-f(u)dt^{2} +\frac{r_{+}^{2} du^{2} }{4u^{3} f(u)} +\frac{r_{+}^{2} }{l^{2}u} d\vec{x}^2 \\
 &=-\frac{r_{+}^{2} }{u\, l^{2} } F(u)dt^{2} +\frac{l^{2} du^{2} }{4u^{2} F(u)} +\frac{r_{+}^{2} }{l^{2}u} d\vec{x}^2 \\
 &\equiv g_{uu} du^{2} +g_{\mu \nu } dx^{\mu } dx^{\nu }
\end{split}
\end{equation}
with $\mu ,\nu =0,..,3$ and $F(u)$  as below,
\begin{equation} \label{eq24}
F(u)=[1-u^{2} +\frac{ul^{2} }{r_{+}^{2} } -\frac{u^{2} l^{2} }{r_{+}^{2} } +\frac{e^{2} u^{2} l^{2} \ln u}{r_{+}^{4} } ] .
\end{equation}

We use the metric (\ref{eq23}) for calculating shear viscosity. The background metric can be perturbed as $g_{\mu\nu} \to g_{\mu\nu} + h_{\mu\nu}$ \cite{Ref15,Ref16,Ref17}. Considering the abbreviation $h_{\mu\nu}\equiv \phi$, the mode equation is found to be,
\begin{equation} \label{eq26}
\frac{1}{\sqrt{-g} } \partial _{u} (\sqrt{-g} g^{uu} \partial _{u} \phi (t,u,\vec{x}) )+g^{\mu \nu } \partial_{\mu } \partial_{\nu } \phi (t,u,\vec{x}) =0,
\end{equation}

Now we apply Fourier transformation from ($t, \vec{x}$) to $k^\mu=(\omega, \vec{k})$ in Eq.(\ref{eq26}). Then ignoring the spatial momentum for simplicity, that is setting $\vec{k}=0 $ in the Green-Kubo formula, we have,
\begin{equation} \label{eq27}
\frac{1}{\sqrt{-g} } \partial _{u} (\sqrt{-g} g^{uu} \partial _{u} \phi )-g^{tt} \omega ^{2} \phi =0.
\end{equation}

Then introducing $\phi=G(u) \phi_{0}(t,\vec{x})$ where $\phi_{0}(t,\vec{x})$ is the source for both graviton in the bulk and the stress tensor on the boundary, we will get,
\begin{equation} \label{eq28}
\frac{d^{2} G(u)}{du^{2} } +(\frac{F'(u)}{F(u)} -\frac{1}{u} )\frac{dG(u)}{du} +\frac{l^{4} \omega ^{2} }{4\, u\, r_{+}^{2} \, F(u)^{2} } G(u)=0,
\end{equation}
with $F'(u) \equiv \frac{d}{du} F(u)$.

\noindent The long wavelength dynamics of strongly coupled field at boundary can be described in terms of the near horizon data of the black brane solution in the bulk space-time. Therefore, we solve the mode equation close to the horizon using the following approximation in the near horizon limit,
\begin{equation} \label{eq30}
\ln(u)|_{u\approx 1} \approx u-1  \nonumber
\end{equation}
\begin{equation} \label{eq31}
F(u)\approx(1-u)[1+u+\frac{ul^{2}}{r_{+}^{2}} -\frac{e^{2} u^{2} l^{2}}{r_{+}^{4}}].
\end{equation}

\noindent Substituting Eq.(\ref{eq31}) into the mode equation (\ref{eq28}) gives us,
\begin{equation} \label{eq32}
\frac{d^{2} }{du^{2} } G(u)-\frac{1}{1-u} \frac{d}{du} G(u)+\frac{\omega ^{2} l^{4} }{4r_{+}^{2} (1-u)^{2} (2+\frac{l^{2} }{r_{+}^{2} } -\frac{e^{2} l^{2} }{r_{+}^{4} } )^{2} } G(u)=0.
\end{equation}

The above equation has a solution in the form of $G(u)=(1-u)^{\beta }$. By putting this ansatz into the  Eq.(\ref{eq32}) we can obtain $\beta$,
\begin{equation} \label{eq33}
\beta =\pm \frac{I\omega \, l^{2} r_{+}^{3} }{2\, (2r_{+}^{4} +l^{2} r_{+}^{2} -e^{2} l^{2} )}=\frac{\pm I\omega}{4\pi T}=\pm \frac{I\varpi }{2} ,\, \, \, \, \varpi =\frac{\omega }{2\pi T}
\end{equation}
where \textit{T} is the Hawking temperature.

\indent Retarded Green's function on the boundary corresponds to the ingoing mode of near horizon. Due to event horizon properties the outgoing mode doesn't exist. Putting the outgoing solution aside we consider the following ansatz for the mode equation Eq.(\ref{eq28}),
\begin{equation} \label{eq35}
G(u)=F(u)^{\frac{-I\varpi }{2} } (h_{0} (u)+\frac{I\varpi }{2} h_1(u)+O(\varpi ^{2} )).
\end{equation}

\noindent Since we want to normalize $G(u)$ on the boundary, we choose $h_{0} (u)=1$. So that,
\begin{equation} \label{eq36}
G(u)=F(u)^{\frac{-I\varpi }{2} } (1+\frac{I\varpi }{2} h_1(u)+O(\varpi ^{2} )).
\end{equation}
Substituting in Eq. (\ref{eq28}), we find for the first order of $\varpi$,
\begin{equation}
h_1''+\left(\frac{F'}{F}-\frac{1}{u} \right) h'_1 -\frac{F''}{F} +\frac{1}{u}\frac{F'}{F} =0
\end{equation}
This is equivalent to
\begin{equation}
\left(\frac{F}{u}h'_1 -\frac{F'}{u} \right)' =0
\end{equation}
Thus
\begin{eqnarray} \label{333}
\frac{F}{u}h'_1 -\frac{F'}{u} = C_1 \\
h_1 = \log \frac{F}{C_2}+C_1\int^u \frac{u}{F}du
\end{eqnarray}
where $C_1$ and $C_2$ are integration constants. The integration in the last line is not elementary. However, by investigating the near horizon behavior, we can determine $C_1$ such that $h_1$ has a regular form at the horizon $u=1$. Using the approximation in (\ref{eq31}) one finds,
\begin{eqnarray}
h_1 &\approx & \log \frac{(1-u)[1+u+\frac{ul^{2}}{r_{+}^{2}} -\frac{e^{2} u^{2} l^{2}}{r_{+}^{4}}]}{C_2}
 - \frac{C_1r_+}{4\pi T  l^2 } \frac{(3+\frac{ l^2}{r_+^2})}{B} \log \frac{B-1+2 \frac{e^2 l^2}{r_+^4}u-\frac{l^2}{r_+^2}}{B+1-2 \frac{e^2 l^2}{r_+^4}u+\frac{ l^2}{r_+^2}}
\nonumber\\
& & -\frac{C_1r_+}{4\pi T  l^2 }\log \frac{(1-u)^2}{-1-u-\frac{u l^2}{r_+^2}+\frac{u^2e^2 l^2}{r_+^4}}
\end{eqnarray}
in which
$$
B=\sqrt{1+6\frac{l^2}{r_+^2}+\frac{l^4}{r_+^4}+4\frac{e^2 l^2}{r_+^4}}
$$
Now taking
\begin{equation}
C_1= \frac{2\pi l^2}{r_+}T
\end{equation}
guarantees that $h_1$ is regular at $u=1$.

The prescription for calculation of retarded Green's function is presented by Son \cite{Ref18,Ref19}. We calculate retarded Green's function by this prescription as follows:
\begin{eqnarray} \label{eq37}
G_{y y} ^{x x} (\omega ,\vec{0})&=&-\sqrt{-g} g^{uu} \, G^{*} (u)\, \partial _{u} G(u)|_{u \to 0} \nonumber\\
 &=& \frac{I r_+^4 \varpi}{\ell^5} \left[ \frac{F'}{u}-\frac{F h'}{u}\right]|_{u \to 0}
  \nonumber\\
 &=& -\frac{I\, \omega \, r_{+}^{4} }{2\pi \, T l^{5} } C_1=-\frac{I\, \omega \, r_{+}^{3} }{l^{3}}
\end{eqnarray}
where in the last line Eq.(\ref{333}) was used.

\noindent Now we can calculate shear viscosity by using Green-Kubo formula \cite{Ref17},
\begin{equation} \label{eq42}
\eta =-\mathop{\lim }\limits_{\omega \to 0} \frac{1}{\omega } \Im G_{y y} ^{x x} (\omega ,\vec{0})=\,\frac{r_{+}^{3}}{l^{3}}
\end{equation}

\noindent Then the ratio of shear viscosity to entropy density is,
\begin{equation} \label{etapers}
\frac{\eta }{s} =\frac{1}{4\pi } .
\end{equation}

\noindent This exactly saturates the conjectured bound and agrees with \cite{Ref17}.


\section{Membrane Paradigm Method}
\label{sec3}
\indent Now we calculate shear viscosity to entropy density via membrane paradigm method. Consider the metric as follow
\begin{equation}
ds^{2}=G_{00}(r)dt^{2} +G_{rr}(r)dr^{2}+G_{xx}(r)\sum_{i=1}^{p}(dx^{i})^{2}+ Z(r)K_{mn}(y)d{y}^{m}d{y}^{n}
\end{equation}
The formula for $\frac{\eta}{s} $ was summarized by \cite{Ref19} as
\begin{equation}\label{eq43}
\frac{\eta }{s} =T\frac{\sqrt{-G(r_{+})}}{\sqrt{-G_{00}(r_{+})G_{rr}(r_{+})}}\int_{r_{+}}^{\infty}{\frac{-G_{00}(r)G_{rr}(r)}{G_{xx}(r)\sqrt{-G(r)}}} dr.
\end{equation}
By considering the metric Eq.(\ref{eq09}) and applying Eq.(\ref{eq43}) we have,
 \begin{equation}
 \frac{\eta }{s} =Tl^{2}r_{+}^{3}\int_{r_{+}}^{\infty}\frac{dr}{r^{5}}=\frac{Tl^{2}}{4r_{+}}=\frac{1}{4\pi}[1+\frac{l^{2}(r_{+}^{2} - e^{2}) }{2r_{+}^{4}}]
 \end{equation}
 In hydrodynamics limit we should consider large $ r_{+} $,
  \begin{equation}\label{mem-para}
   \frac{\eta }{s}\approx \frac{1}{4\pi}.
   \end{equation}

Comparing Eq. (\ref{mem-para})  with the Green-Kubo result in Eq. (\ref{etapers}) shows that two results agree in hydrodynamics regime or the large $r_+$ limit.

Of course there is another route to calculate the ratio in the context of membrane paradigm \cite{Iqbal:2008by}. This gives exactly $\frac{1}{4\pi}$ and its approach is closer in spirit to the Kubo formalism.

 \section{Conclusion}

\noindent We showed that the lower bound of the ratio $\eta/s$ preserves for EYM black brane. This bound is known as KSS conjecture \cite{Ref17} and considered for strongly interacting systems where reliable theoretical estimate of the viscosity is not available. It tells us that the ratio $\eta/s$ has a lower bound, $\frac{\eta }{s} \ge \frac{\hbar }{4\, \pi \, k_{B} } $, for all relativistic quantum field theories at finite temperature without chemical potential \cite{Ref4,Ref15,Ref17} and can be interpreted as the Heisenberg uncertainty principle \cite{Ref15}.  

However it is well-known that $\eta/s$ for those theories with an Einstein gravity dual saturates the bound (i.e. $\eta/s=1/4\pi$). This was firstly pointed out in \cite{Ref17} by noticing that the metric perturbation which was used to find the viscosity on the boundary obeys a minimally coupled massless scalar equation. Thus the viscosity is the same as the scalar absorption cross section which in turn, at low frequencies, is found to be the horizon area. On the other hand the horizon area is proportional to the entropy, so the viscosity to entropy ratio is found to be  constant $1/4\pi$. Of course this result gets modification and  the bound is violated in the case of higher derivative gravities like the Gauss-Bonnet gravity \cite{Ref22,Ref23,Ref24}.

In this article, the ratio $\eta/s$ was calculated from two different approaches, the Green-Kubo formula and the membrane paradigm integral formula. The former gives $1/4\pi$ exactly, while the latter reaches $1/4\pi$ in the large $r_+$ limit. This limit corresponds to large black holes where the hydrodynamic limit is valid for the corresponding field theory. It is worth to comment on this discrepancy between two approaches. As explained in \cite{Buchel:2003tz} the underlying symmetry for deriving the universal result of $\eta/s=1/4\pi$ is the Poincare invariance of the background which can be implemented by the equation $R_{tt}+R_{xx}=0$. In our case, this condition is not satisfied by (\ref{eq09}). Indeed we used small solid angle or large $r_+$ radius approximation to convert the black hole in (\ref{eq09}) into a membrane with Poincare invariance as needed for a hydrodynamic description. The curios point is that the Kubo formula does not depend on the details of the black hole or black brane solution while the membrane paradigm integral formula, as our analysis reveals, is sensitive to the details of the solution. As mentioned before there is an alternative membrane paradigm method given in \cite{Iqbal:2008by} by which  $\eta/s=1/4\pi$ result can be achieved exactly. In this latter method, instead of taking an integral from horizon to the boundary, viscosity can be found by geometric properties at the horizon and in some sense is in spirit of the Kubo formula with the same exact result \footnote{See also \cite{Sasai:2010pz} for viscosity to entropy ratio of black holes membrane paradigm in the context of highly excited string.}.

\vspace{1cm}
\noindent {\large {\bf Acknowledgment} } Authors would like to thank Komeil Babaei, Ali Imaanpur and Viktor Jahnke for useful discussions and comments.

\vspace{1cm}



\begin{thebibliography}{}

\bibitem{Ref1}
   J. M. Maldacena, ``The Large N limit of superconformal field theories and supergravity,'' Int.\ J.\ Theor.\ Phys.\  {\bf 38} (1999) 1113 [Adv.\ Theor.\ Math.\ Phys.\  {\bf 2} (1998) 231] [hep-th/9711200].

\bibitem{Ref2}
 O. Aharony, S.~S.~Gubser, J.~M.~Maldacena, H.~Ooguri and Y.~Oz,
 ``Large N field theories, string theory and gravity,''
  Phys.\ Rept.\  {\bf 323}, 183 (2000)
  [hep-th/9905111].

\bibitem{Ref3}
 J.~Casalderrey-Solana, H.~Liu, D.~Mateos, K.~Rajagopal and U.~A.~Wiedemann,
      ``Gauge/String Duality, Hot QCD and Heavy Ion Collisions,''  arXiv:1101.0618 [hep-th].

\bibitem{Ref4}
D.~Mateos,
  ``String Theory and Quantum Chromodynamics,''
  Class.\ Quant.\ Grav.\  {\bf 24}, S713 (2007)
  [arXiv:0709.1523 [hep-th]].

\bibitem{Ref5}
S.~Bhattacharyya, V.~E.~Hubeny, S.~Minwalla and M.~Rangamani,
 ``Nonlinear Fluid Dynamics from Gravity,''
  JHEP {\bf 0802}, 045 (2008)
  [arXiv:0712.2456 [hep-th]].

\bibitem{Ref6}
  N.~Ambrosetti, J.~Charbonneau and S.~Weinfurtner,
   ``The Fluid/gravity correspondence: Lectures notes from the 2008 Summer School on Particles, Fields, and Strings,''
   arXiv:0810.2631 [gr-qc].

\bibitem{Ref7}
M.~Rangamani,``Gravity and Hydrodynamics: Lectures on the fluid-gravity correspondence,''
  Class.\ Quant.\ Grav.\  {\bf 26}, 224003 (2009)
  [arXiv:0905.4352 [hep-th]].

\bibitem{Ref8}
L. D. Landau and E. M. Lifshitz, Fluid Mechanics (Course of Theoretical Physics,Vol. 6). Butterworth-Heinemann, 1965.

\bibitem{Ref9}
 J.~Bhattacharya, S.~Bhattacharyya, S.~Minwalla and A.~Yarom,
   ``A Theory of first order dissipative superfluid dynamics,''
   JHEP {\bf 1405}, 147 (2014)
   [arXiv:1105.3733 [hep-th]].

\bibitem{Ref10}
 P. Kovtun,``Lectures on hydrodynamic fluctuations in relativistic theories,''J.\ Phys.\ A {\bf 45} (2012) 473001[arXiv:1205.5040 [hep-th]].

\bibitem{Ref11}
 N.~Banerjee and S.~Dutta,
  ``Holographic Hydrodynamics: Models and Methods,''
  arXiv:1112.5345 [hep-th].

\bibitem{Ref12}
D.T. Son,  Hydrodynamics and gauge/gravity duality, Nuclear Physics B (Proc. Suppl.) 192--193 (2009) 113--118.

\bibitem{Ref13}
 Joseph I. Kapusta , Charles Gales; ``finite- temperature field Theory Principle and applications'' cambridge university press 2006.

\bibitem{Ref14}
 N.~Bostani and M.~H.~Dehghani,
   ``Topological Black Holes of (n+1)-dimensional Einstein-Yang-Mills Gravity,''
   Mod.\ Phys.\ Lett.\ A {\bf 25}, 1507 (2010)
   [arXiv:0908.0661 [gr-qc]].

\bibitem{Ref15}
 D.~T.~Son and A.~O.~Starinets,
  ``Viscosity, Black Holes, and Quantum Field Theory,''
  Ann.\ Rev.\ Nucl.\ Part.\ Sci.\  {\bf 57}, 95 (2007)
  [arXiv:0704.0240 [hep-th]].

\bibitem{Ref16}
 G.~Policastro, D.~T.~Son and A.~O.~Starinets,
   ``The Shear viscosity of strongly coupled $ \mathcal{N}=4 $ supersymmetric Yang-Mills plasma,''
   Phys.\ Rev.\ Lett.\  {\bf 87}, 081601 (2001)
   [hep-th/0104066].

\bibitem{Ref17}
 P.~Kovtun, D.~T.~Son and A.~O.~Starinets,
    ``Viscosity in strongly interacting quantum field theories from black hole physics,''
    Phys.\ Rev.\ Lett.\  {\bf 94}, 111601 (2005)
    [hep-th/0405231].

\bibitem{Ref18}
G.~Policastro, D.~T.~Son and A.~O.~Starinets,``From AdS/CFT correspondence to hydrodynamics,''
  JHEP {\bf 0209}, 043 (2002)
  [hep-th/0205052].

\bibitem{Ref19}
 P.~Kovtun, D.~T.~Son and A.~O.~Starinets,
    ``Holography and hydrodynamics: Diffusion on stretched horizons,''
    JHEP {\bf 0310}, 064 (2003)
    [hep-th/0309213].

\bibitem{Ref20}
 U.~Gürsoy, I.~Iatrakis, E.~Kiritsis, F.~Nitti and A.~O'Bannon,
   ``The Chern-Simons Diffusion Rate in Improved Holographic QCD,''  JHEP {\bf 1302}, 119 (2013)  [arXiv:1212.3894 [hep-th]].

\bibitem{Ref21}
 V.~Jahnke, A.~S.~Misobuchi and D.~Trancanelli,
  ``The Chern-Simons diffusion rate from higher curvature gravity,''  Phys.\ Rev.\ D {\bf 89}, 107901 (2014)  [arXiv:1403.2681 [hep-th]].


\bibitem{Ref22}
M.~Brigante, H.~Liu, R.~C.~Myers, S.~Shenker and S.~Yaida,
  ``Viscosity Bound Violation in Higher Derivative Gravity,''
  Phys.\ Rev.\ D {\bf 77}, 126006 (2008)
  [arXiv:0712.0805 [hep-th]].


\bibitem{Ref23}
M.~Brigante, H.~Liu, R.~C.~Myers, S.~Shenker and S.~Yaida,
   ``The Viscosity Bound and Causality Violation,''
   Phys.\ Rev.\ Lett.\  {\bf 100}, 191601 (2008)
   [arXiv:0802.3318 [hep-th]].

 \bibitem{Ref24}
  I.~P.~Neupane and N.~Dadhich,
     ``Entropy Bound and Causality Violation in Higher Curvature Gravity,''
     Class.\ Quant.\ Grav.\  {\bf 26}, 015013 (2009)
     [arXiv:0808.1919 [hep-th]].

\bibitem{Iqbal:2008by} 
  N.~Iqbal and H.~Liu,
  Phys.\ Rev.\ D {\bf 79}, 025023 (2009)
  [arXiv:0809.3808 [hep-th]].

\bibitem{Buchel:2003tz} 
  A.~Buchel and J.~T.~Liu,
  Phys.\ Rev.\ Lett.\  {\bf 93}, 090602 (2004)
  [hep-th/0311175].
  
\bibitem{Sasai:2010pz} 
  Y.~Sasai and A.~Zahabi,
  Phys.\ Rev.\ D {\bf 83}, 026002 (2011)
  [arXiv:1010.5380 [hep-th]].
  
\end{thebibliography}
\end{document}